# CMOS based high-resolution dynamic X-ray imaging with inorganic perovskite

Yanliang Liu[a], Chaosong Gao[b], Jiongtao Zhu[a], Xin Zhang[a], Meng Wu[b], Ting Su[a], Jiahong Wang[a], Zonghai Sheng[a], Wenjun Liu[a], Tongyu Shi[a], Xingchen He[a], Dong Liang[a], Hairong Zheng[a], Xue-Feng Yu[a], Xiangming Sun[b], Yongshuai Ge[a]

[a]Shenzhen Institute of Advanced Technology, Chinese Academy of Sciences, China

[b]Central China Normal University, Wuhan, Hubei, China

*Y.Liu and C.Gao made equal contribution to this work.*
*Email: xf.yu@siat.ac.cn, xmsun@phy.ccnu.edu.cn, ys.ge@siat.ac.cn*

**Abstract:** High-resolution dynamic X-ray detector is crucial for time-resolved digital radiography (DR) imaging and fast 3D medical computed tomography (CT) imaging. Recently, perovskites have become promising alternatives to conventional semi-conductor materials, e.g., Si, a-Se and CdTe, for direct X-ray detection. However, the feasibility of their combination with high-speed pixelated complementary metal-oxide-semi- conductor (CMOS) arrays remains unknown. This work originally reports an innovative direct-conversion X-ray detector fabricated with 300 micrometer thick inorganic perovskite film printed on a tailored CMOS array. In-house measurements demonstrate that the $CsPbBr_3$ film has excellent optoelectric properties of a μτ product of $3.40 \times 10^{-5}$ $cm^2$ $V^{-1}$, and the X-ray detector exhibits high sensitivity of 9341 μC $Gy_{air}^{-1}$ $cm^{-2}$, and low detection limit of 588 $nGy_{air}$ $s^{-1}$. This CMOS X-ray imaging detector achieves a high spatial resolution up to 5.5 lp/mm (close to the resolution limit of 6.0 lp/mm), and >300 frame per second (fps) readout speed. DR image of a resolution pattern phantom and a anesthesia mice, CT images of a biological specimen are acquired for the first time.

Since its discovery in 1895, X-ray imaging has shown superb performance in unveiling the anatomical sturctures of the human body. By far, X-ray imaging has been widely used in many medical applications, such as the diagnosis and treatment of cardiovascular and cancer diseases[1,2]. In order to generate sufficient X-ray imaging information for clinicians to make precise diagnoses, usually, X-ray detectors with unique features such as high spatial resolution, fast imaging speed are highly demanded. In addition, to minimize its cancerous damage to human body, its low dose X-ray imaging performance is also an important factor to be considered to use as low as reasonably achievable (ALARA) radiation dose[3].

Over the past two decades, dedicated academic and clinical investigations have been performed to develop the most appropriate X-ray detectors for ultra-high quality medical imaging. Results demonstrate the direct-conversion type X-ray detectors made of semi-conductor materials show better spatial and temporal resolution at low radiation dose X-ray imaging than the indirect-conversion ones made of scintillator materials[4]. However, the current semi-conductor materials used in commercial direct-conversion type X-ray detectors are less satisfactory for generic X-ray imaging purposes. For example, the amorphous selenium (a-Se)[5,6] only works at low energy X-ray photons (<40 keV) due to its low stopping power (Z=34), which are merely restricted for breast imaging. Moreover, the cadmium zinc telluride (CdZnTe)[7] or cadmium telluride (CdTe) crystals have shown its advancements in fabricating high-energy (>140 keV) X-ray detectors with energy-resolving capability. However, the CdZnTe/CdTe are are difficult to grow into large dimensions, and thus are hard to be used for large area X-ray imaging. The high cost also prohibits their wide spreads in medical X-ray imaging applications. In addition, the requirement of flip chip assembly using Indium solder bumps to interconnect the CdZnTe/CdTe crystal slab to the back-end ASIC readout circuit complicates the detector fabrication as well.

As an emerging candidate, metal halide perovskites[8,9,10,11], e.g., single crystal, polycrystalline wafer and solvent-processed thick film, show excellent potentials in realizing high-sensitive (≥50000 µC $Gy_{air}^{-1}cm^{-2}$) direct X-ray detection, such as high X-ray absorption, high charge carrier mobility (µ) and long carrier lifetime (τ). By integrating with pixelated sensor arrays, prototypes of perovskite-based X-ray imaging detectors have been successfully fabricated. In 2017, Park *et.al.* first reported the promise of high-resolution X-ray imaging with blade-coating $CH_3NH_3PbI_3$ films on a large-scale thin-film transistor (TFT) backplane[12]. In 2021, Sarah *et.al.* reported a two-step procedure to manufacture X-ray detector with microcrystalline $CH_3NH_3PbI_3$ on TFT backplane[13]. Recently, Tang *et.al.* fabricated a flat-panel X-ray imager by combination of soft-presssing $CH_3NH_3PbI_3$ films and TFT backplane[14].

Technically, the direct-conversion perovksite X-ray detectors having the most ideal imaging performance should be fabricated upon the Complementary Metal-Oxide-Semiconductor (CMOS) sensors[15], which have become the majority of consumer and prosumer cameras these days due to their much more distinct advantages[16,17,18] in achieving ultra-high image resolution and data readout speed[19]. For example, the pixel size of CMOS sensor can be easily made smaller than 5 micrometers, whereas, the pixel size of TFT array is usually larger than 70 micrometers. Besides, the CMOS sensor also has stronger noise immunity and lower static power utilization. As a consequence, CMOS sensor arrays should be selected to completely fulfill the advancements of perovksites in achieving high-end direct-conversion X-ray imaging with unprecedented spatial and temporal resolution. Moreover, special electrical circuits can be designed and integrated into the CMOS pixel unit to surpress the leakage dark current of the perovksites[20], which is still a remaining issue in developing X-ray imaging detector array[21]. Unfortunately, by far no such studies have been demonstrated on a CMOS sensor based perovskite X-ray detector.

In this work, we report an innovative direct-conversion X-ray detector fabricated with 300 micrometer thick inorganic $CsPbBr_3$ perovskite film printed on a tailored CMOS sensor array[22,23]. The main structure of the X-ray imager was illustrated in Fig. 1a. As shown, certain electric field is applied to propel the X-ray photon stimulated electrons within the perovskite to drift towards the signal collection electrode, namely, the CMOS pixel. The perovskite film covered CMOS device with total number of 72×72 pixels is depicted in Fig. 1b. The perovskite film was printed directly on the CMOS chip via the silk-screen priting process. The cross-section SEM image shows that such in-situ grown $CsPbBr_3$ film has good affinity with the CMOS array, in which multi-layered parasitic capacitance with multifunctional complex circuits of amplifier and leakage current compensation are integrated to enhance high SNR signal detections. The detector could endure more than four months without significantly degrading its X-ray imaging performance, indicating a excellent working stability of $CsPbBr_3$ film under X-ray exposure and high electric field conditions. This CMOS based direct-conversion X-ray detector can resolve objects with a very fine structure of 5.5 lp/mm, see Fig. 1c. It was obtained at a bias voltage of 15V, corresponding to an electric field of 0.05 V/um inside the perovskite film. This experiment was validated by a bar pattern plate having 33 groups of different Pb lines with varied widths and spacing distances. Since the limit spatial resolution of this CMOS array is found to be 6.0 lp/mm (corresponds to the 83.2 μm pitch dimension), therefore, this X-ray detector shows an unprecedented high-resolution imaging performance. For this special CMOS array, the gate voltage of the feedback transistor, denoted as V, controls the decay time of the Charge Sensitive Amplifier (CSA) for every single pixel (see Methods for details). To inhibit the pixel-by-pixel varied leakage current, especially to avoid the signal saturations, the gate voltage was adjusted according to the amplitude of the leakage current without turning on the X-ray beam. The 3D distribution map of the pixel response is illustrated in Fig. 1d. In general, the pixel response decreases as the used gate voltage increases. To suppress the internal dark current of the perovskite film, certain gate voltages are calibrated and adopted before X-ray exposures. For larger dark current signals, higher gate voltages are needed to maintain similar pixel responses.

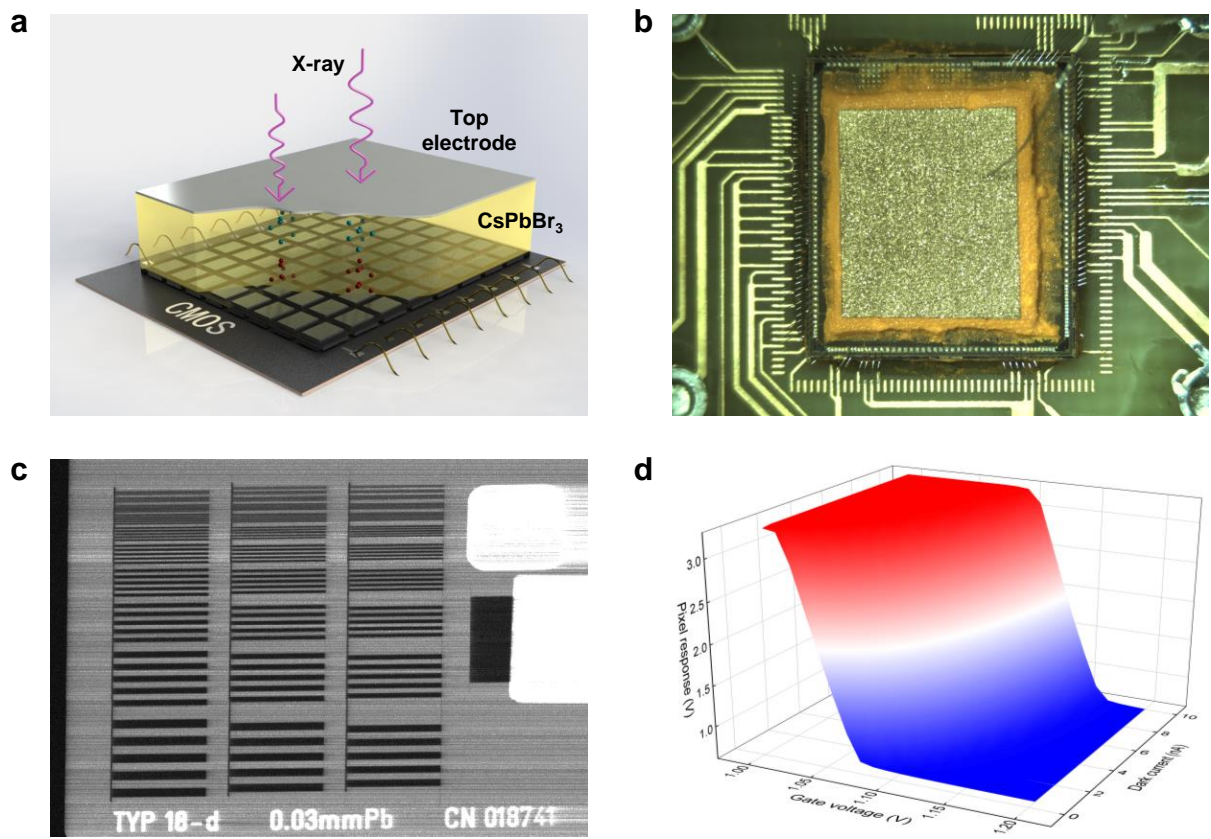

**Fig 1. | CMOS based perovskite X-ray detector. a,** diagram of the X-ray imaging panel based on CMOS array. Electron-hole pairs are stimulated by the X-ray photons within the perovskite film. The electrons drift swiftly towards the CMOS pixel under the internal electric field. **b,** photograph of the CMOS based perovskite X-ray detector sealed inside a costumized Teflon holder. Desiccant beads are added to keep the detector away from moist air. **c,** the obtained X-ray image of a resolution pattern phantom. The bars with 5.5 lp/mm spatial resolution can be clearly distinguished. **d,** 3D response map of a single CMOS pixel versus the added gate voltage and the input dark current. The pixel response decreases as the gate voltage increases. For larger dark current signals, higher gate voltages are needed.

The inorganic $CsPbBr_3$ perovskite thick film was fabricated through silk-screen priting process, see Fig.2a. First, a viscous $CsPbBr_3$ precursor paste was prepared by mixing CsBr and $PbBr_2$ with equal molar mass in DMF/DMSO primary polar solvent through planetary ball milling. Second, the silk-screen prited CsPbBr$_3$ precursor thick film was heated at 100 °C for 10 min to obtain an intermediate CsPbBr$_3$ thick film, which still contains trace amount of residual DMSO according to the Fourier-transform infrared (FTIR) spectra. Afterwards, the intermediate CsPbBr$_3$ film was hot-pressed at 150 °C and 0.5 MPa through a self-desioned equipment.The hot-pressed CsPbBr$_3$ thich film owns flat surface and compact internal structure in contrast to conventional CsPbBr$_3$ film with rough surface and pinholes inFig.2b,c. In general, the hot-presseded CsPbBr$_3$ thick film with compact and uniform surface has better capability to inhibit leak current, ionic migration, photo-induced carrier trapping, and recombination than the porous film that is produced without hot-pressing. Moreover, the heat-treated CsPbBr$_3$ thick film shows significantly increased XRD intensity at the main (001), (011) and (002) peaks, suggesting better crystallinity in line with the SEM images in Fig.2d. Be aware that the thickness of CsPbBr$_3$ film can be easily adjusted from tens to hundreds micrometers if repeating the silk-screen priting procedures. The CsPbBr$_3$ film was X-rayed themselves, and their X-ray absorption can be adjusted by controlling the CsPbBr$_3$ film thickness. For X-ray imaging, large-area fabrication of perovskite film is important, we therefore prepared a large area CsPbBr$_3$ film of 10×10 cm by silk-screen printing on a ITO/glass substrate, the film exhibits compact and uniform morphology. In addition, the obtained CsPbBr$_3$ film has excellent environmental stability with stable optical property and crystal structure, which is benefitial for enhancing the working stability of the X-ray detector.

The steady-state photoluminescence (PL) and time-resolved photoluminescence (TRPL) behaviors of the CsPbBr$_3$ before and after the heating procedures were characterized with the purpose to elucidate their optical property (Fig.1e and Extended Data Fig. S). The treated CsPbBr$_3$ film shows significantly higher PL intensity with PL peak red-shifted from 515 nm to 518 nm, indicating the reduced defects density and enhanced crystallinity. The measured TRPL values were fitted with exponential decay functions to measure the quenching (including charge transfer) and defects related fast decay process, i.e., short lifetime ($\tau_1$), the recombination related slow decay process, i.e., long lifetime ($\tau_2$), and their occupation fractions $A_1$ and $A_2$. The CsPbBr$_3$ before treating has a short lifetime $\tau_1$ =2.29 ns and a long lifetime $\tau_2$ =11.74 ns, the CsPbBr$_3$ after treating has a short lifetime $\tau_1$ = 5.70 ns and a long lifetime $\tau_2$ = 33.07 ns, and the $\tau_{ave}$ improved significantly from 6.80 to 25.92 ns, indicating reduced defect density and enhanced carrier transportation capability after treatment. The µτ product of the treated CsPbBr$_3$ thick film is obtained by fitting the following Hecht equation from the measured voltage-dependent photocurrents:

$$I=I_0\mu\tau V/L^2(1-\exp(-L^2/\mu\tau V)),$$

where $I_0$ is the saturated current, L is the MAPbI$_3$ wafer thickness, and V is the bias voltage. The µτ product of our CsPbBr$_3$ film is determined to be 3.40×10$^{-5}$ cm$^2$ V$^{-1}$, which is significantly higher than the amorphous selenium and thus can benefit the X-ray detection.

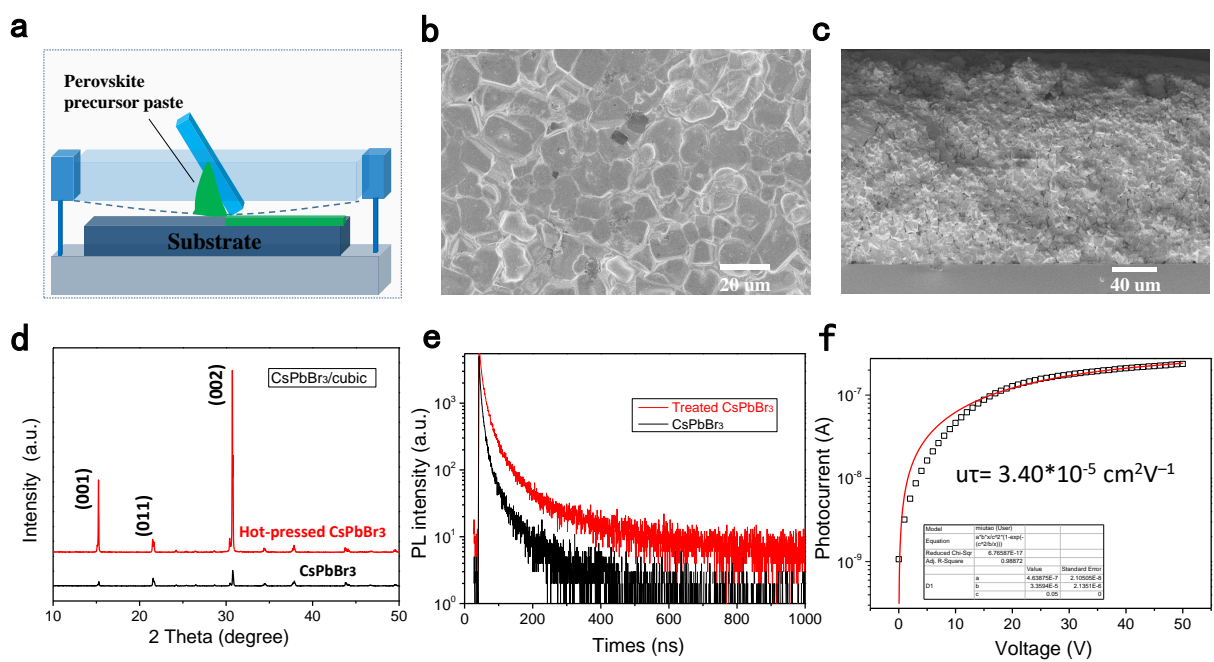

**Fig 2. | Inorganic CsPbBr3 perovskite thick film.** (a) Schematic illustration of the silk-screen priting CsPbBr3 film. (b) top-view and (c) cross-section SEM images of CsPbBr3 film. (d) XRD

spectra of CsPbBr3 film. (e) TRPL spectra of CsPbBr3 film. (f) photoconductivity of the CsPbBr3 film.

The experimental current density-voltage (J-V) curves of the CsPbBr$_3$ thick film are shown in Fig.3a. From the top to the bottom, the device contains glass-ITO/SnO$_2$/CsPbBr$_3$/ Au. The dark current density is as low as $1.41\times10^{-9}$ A mm$^{-2}$ and the corresponding light current is up to $2.10\times10^{-7}$ A mm$^{-2}$ at the electric field of 80 V/mm, resulting a switch ratio of 150. Time-resolved current densities of the CsPbBr$_3$ detector at various X-ray dose rates varied from 160 µGy$_{air}$ s$^{-1}$ to 1620 µGy$_{air}$ s$^{-1}$ are measured, see Fig 3b. As seen, the detector has a linear response to the X-ray dose rates, and owns smaller dark current densities and larger values of X-ray on-off ratios. Different current densities under varied dose rates are shown in Fig. 3c. It is estimated that the CsPbBr$_3$ detector has a sensitivity of 1547, 4334, 7266 and 9341 µCGy$_{air}^{-1}$ cm$^{-2}$ at electric fields of 20, 40, 60 and 80 V/mm, respectively.

The signal-to-noise ratios (SNR), defined as SNR=$I_{signal}$ / $I_{noise}$ = ($I_{photo}$ − $I_{dark}$) / $I_{noise}$, were calculated, see Fig. 4d. Herein, $I_{signal}$ denotes the signal, $I_{noise}$ denotes the noise, $I_{photo}$ denotes the average current under X-ray irradiation, and $I_{dark}$ denotes the average dark current calculated from several parallel experiments at each bias. If assuming the minimum detective SNR=3, the lower limit-of-detection (LoD) of our CsPbBr$_3$ detector is determined to be 588 nGy$_{air}$ s$^{-1}$.

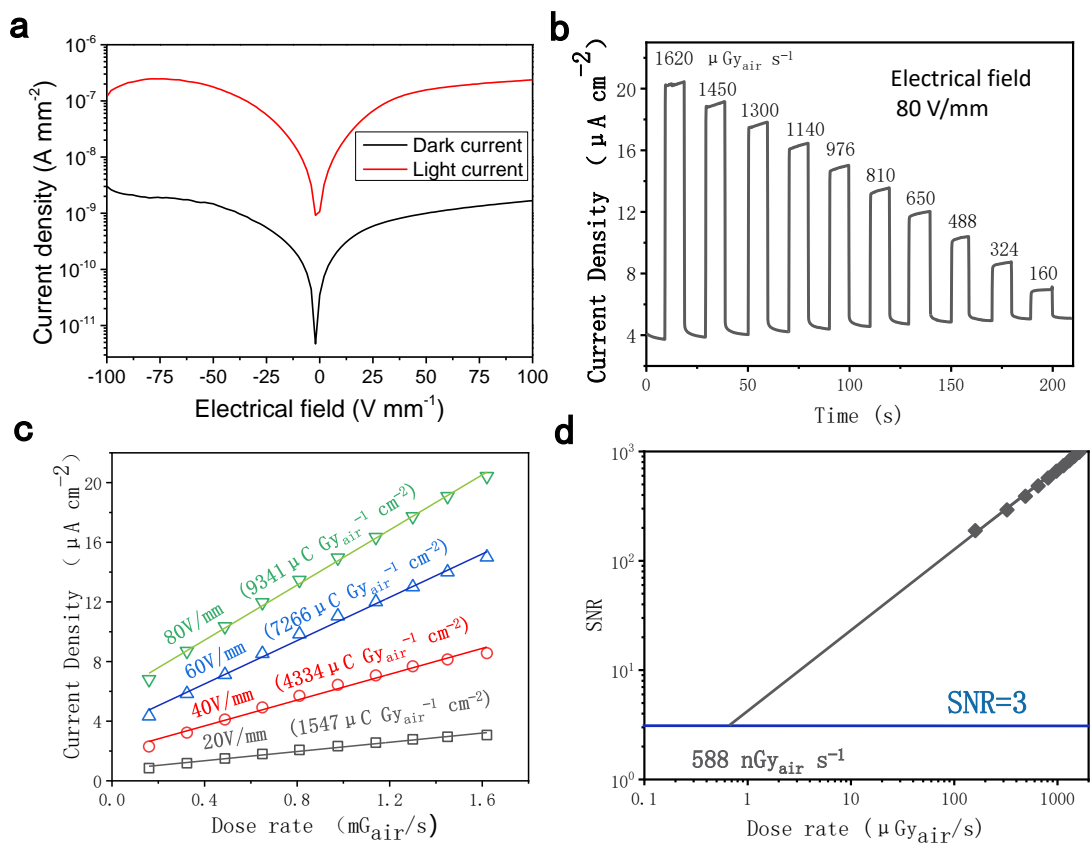

**Fig 3. | Perovskite X-ray detector properties.** (a) J-V characteristic and (b) X-ray response profiles of the CsPbBr3 detector. (c) X-ray current densities as a function of dose rate at different applied electric fields. (d) Dose rate dependent SNR of the CsPbBr3 detector.

The 2D digital radiography (DR) and the 3D computed tomography (CT) imaging results of biological specimen are depicted in Fig. 4, correspondingly. First of all, a three-month old mice after anesthesia was scanned at 50 kVp tube voltage under three different dose levels using this perovskite CMOS detector. To compare, the same mice was also imaged with a commercial scintillator (CsI:Tl) based indirect-conversion CMOS flat panel detector (Model: Dexela 2329 NDT, PerkinElmer, USA) under the same imaging condition. Overall, the DR images obtained from this novel direction-converson perovskite CMOS detector shows superior performance than the conventional indirect-conversion CMOS detector. For example, the ribs and forehand bones can be clearly recognized, see Fig. 4a. Moreover, images obtained from the new detector also look much more crisp. This demonstrates that the perovskite CMOS detector has higher X-ray conversion efficiency, and thus is able to significantly reduce the imaging radiation dose level. Quantitatively, at leaset 70% radiation dose can be saved without dramatically degrading the image quality. Results show that a low detection limit of this CMOS detector in generating X-ray images is less than 50 nGy$_{air}$ (measured with 50 kV X-ray spectra). In Fig. 4c, line profiles of the ribs are compared. As seen, the developed new CMOS detector has better image contrast than the commercial COMS detector. To demonstrate its dynamic imaging capability, 3D CT imaging were performed on a chicken drumstick specimen with this perovskite CMOS detector, see results in Fig. 4b. In total, 180 projections were acquired in a sequence by rotating the sample continuously with two degree angular interval (see Methods for details). After that, CT images were reconstructed via the standard Feldkamp-Davis-Kress (FDK) algorithm with the Ramp filter[24,25]. As seen, the bony structures and the tissue edges can all be clearly recovered. To quantify the spatial resolution performance, the modulation transfer function (MTF) was measured with a 1.0 mm thick tungsten plate[26], one of whose edges was manually tilted by 3.0 degrees relative to the vertical direction. The obtained MTF and normalized noise power spectra (NPS) plots are illustrated in Fig. 4d. In addition, the NPS curves were also characterized[27] from the acquired 200 frames of object-free images (air scans), which have substracted the dark background and been gain corrected via the rounte signal processing procedure. Suprisingly, the NPS of this perovskite CMOS detector is almost flatten, indicating a negilible signal correlations (cross-talking) between neighbouring pixels. In other words, the movements of the stimulated electrons inside the material from X-ray photons are nicely confined by the electric field within a single pixel.

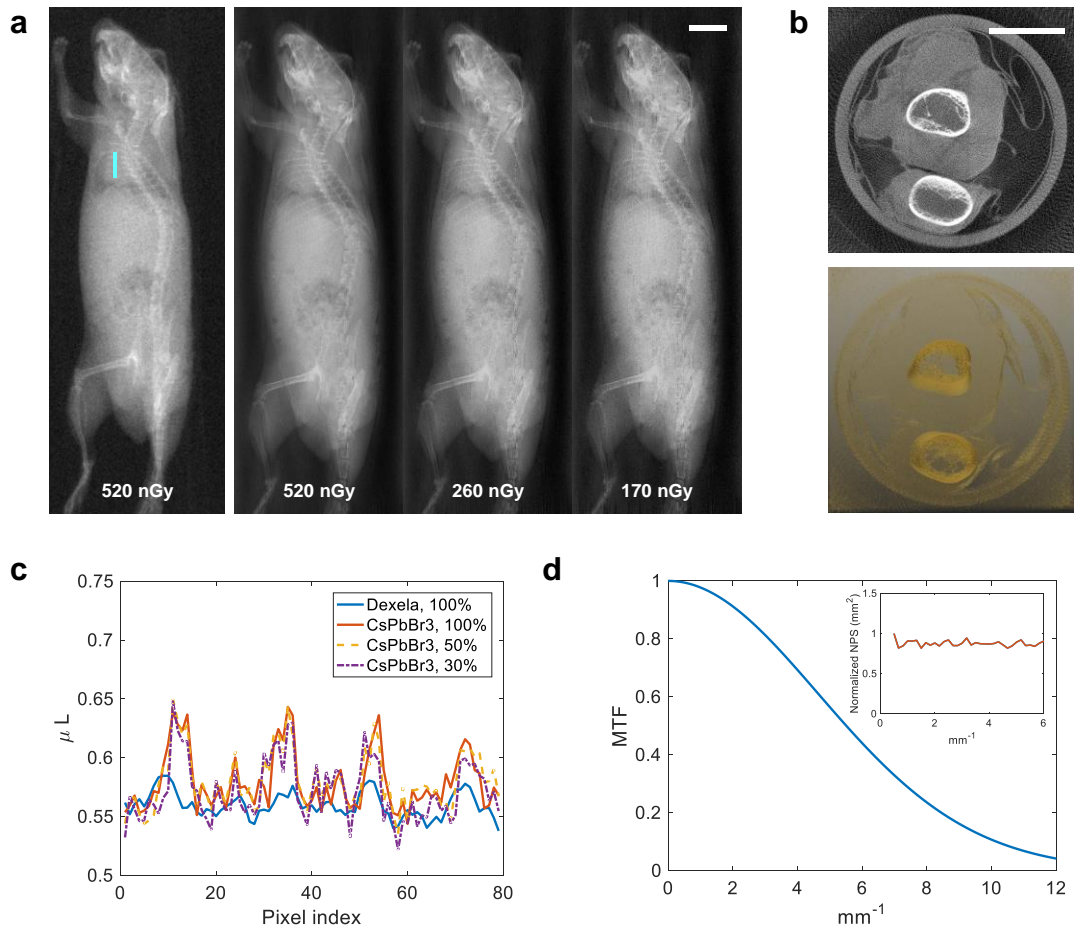

**Fig 4. | Perovskite X-ray imaging results. a,** the DR imaging result of a mice. Image in the first column is obtained from a commercial indirect-conversion CMOS X-ray detector (74.8 um pixel size). Images in the rest three columns are obtained from the new direct-conversion perovskite CMOS X-ray detector (83.2 um pixel size) at three dose levels. The sacle bar denotes 5.0 mm. **b,** the CT imaging results of a chicken drumettes specimen. The upper image shows one single CT image slice, and the lower image shows the 3D volume rendering effect. The sacle bar denotes 10.0 mm. **c,** the line profile comparison results of the highlighted region on the mice. As seen, the developed new CMOS detector has better image contrast than the commercial COMS detector (solid blue line). **d,** MTF and NPS plots of the new direct-conversion perovskite CMOS detector. The former detector was measured at 50 kVp with 1.0 mm aluminum filtration, and the

latter detector was measured with RQA3 beam quality (50 kVp with 10.0 mm aluminum filtration).

In summary, we report an innovative CMOS based direct-conversion high-resolution dynamic X-ray imaging detector fabricated from inorganic perovskite. The 300 micrometer thick $CsPbBr_3$ film was directly printed on the CMOS array via the silk-screen priting process. Experiments demonstrate that the $CsPbBr_3$ film has excellent optoelectric properties of a µτ product of $3.40\times10^{-5}$ $cm^2$ $V^{-1}$, high sensitivity of 9341 µC $Gy_{air}^{-1}$ $cm^{-2}$, and low detection limit of 588 $nGy_{air}$ $s^{-1}$. Both resolution pattern and W-plate imaging results show that this CMOS detector is able to achieve high spatial resolution up to 5.5 lp/mm, which is very close to its resolution limit of 6.0 lp/mm. In addition, signals collected from this CMOS detector are less correlated, and thus leading to an almost flatten NPS curve. Benefited by the >300 fps fast signal readout speed, samples with size larger than 60 mm×60 mm can be quickly scanned. Both DR and CT images are acquired for different biological specimen to demonstrate its high performance 2D and 3D bio-imaging capability.

In conculusion, perovskites provide a completely new blueprint in developing the next generation X-ray imaging detector to reform the conventional bio-medical X-ray imaging applications. As the detector spatial resolution, readout speed and dose conversion efficiency become significantly increased, which have already been demonstrated in this study, the medical X-ray imaging would become more gentle and safe. In the near future, bio-medical X-ray imaging applications such as the small animal imaging, dental imaging and breast imaging would be greatly benefitted if assembling multiple CMOS detector components into a linear detector array or fabricating a large area flat panel from stitched CMOS arrays.

**Supplementary material**

PbBr$_2$ (lead bromide, 99%, Aladdin), CsBr (cesium bromide, 99.5%, Aladdin). Organic solvents including dimethyl formamide (DMSO, AR, 99%) and dimethyl sulfoxide (DMSO, AR, 99%) were purchased from Sigma-Aldrich. The CsBr (6.4 g) and PbBr2 (11.0 g) with equal molar mass were mixed with 6 ml DMF/DMSO (1:1) solvent in agate jar, and then the mixture were grinded in ball-milling for three hours, obtaining a viscous CsPbBr$_3$ precursor paste for silk-screen printing. The susbtrates (glass/ITO or CMOS) were cleaned with detergent, ultrasoni-cated in acetone and isopropyl alcohol, and subsequently dried overnight in an oven at 100 °C. The SnO$_2$ solvent was spin-coated on substrate (4000 rmp, 40 s), then the substrate was annealing at 140 °C for 1 hour, forming a electron transport layer. Then, the viscous CsPbBr$_3$ precursor paste was screen-printed on the substrate, obtaining an CsPbBr$_3$ precursor film, and its thickness can be adjusted by controlling the printing times. The CsPbBr$_3$ precursor film was pre-heated at 100 °C for 10 min, resulting an intermediate CsPbBr$_3$ thick film, which still contains trace amount of residual DMSO. Afterwards, the intermediate CsPbBr$_3$ film was hot-pressed at 150 °C and 0.5 MPa for 10 min through a self-desioned equipment, the hot-pressed CsPbBr$_3$ thich film was further heated at at 150 °C for 30 min, achieving a complete CsPbBr$_3$ thich film. Finally, an Au anode was deposited by thermal evaporation in a vacuum of about 5 × 10$^{-7}$ Torr.

The used CMOS chip consists of a 72×72 active pixel array, a scan module, a switch array and a buffer. The analog signal of each pixel is transmitted off chip pixel by pixel through the switch array and the buffer. The switch array is controlled by the scan module. The charge collection electrode, denoted as T, is an exposed top-most metal. Another top-most metal, denoted as G, covered by an insulation material surrounds around the Topmetal to form a focused electric field hence improving the charge collection efficiency. The detailed circuit of each pixel is composed of a charge collection electrode, a Charge Sensitive Amplifier (CSA) and a two-staged cascaded source follower. The charge collected by the charge collection electrode is fed into the CSA and then converted into a voltage signal. The gate voltage labeled FB_VREF of the feedback transistor can be adjustable to adjust the decay time of the CSA. Hence the leakage current caused by the perovskite thick film can be partially absorbed by increasing the FB_VREF, but the signal is also decreased.

The rear signal readout circuit is composed of a CMOS chip binding board, a mixed signal converter board and a Field-Programmable Gate Array (FPGA) based control board. The binding board carries a bare die of the CMOS chip. Filtering capacitors are placed on the binding board to suppress noise. The binding board is connected with the mixed signal converter board through a pin header. The single-end analog signal output from the CMOS chip is fed into the mixed signal converter board and then converted to a differential signal to match the input of the analog-to-digital converter (ADS5282, TI, USA). The ADS5282 features a digital resolution of 12 bit, a sampling rate of 65 MSps and an interface of the serialized Low Voltage Differential Signaling (LVDS) outputs. A 16-bit digital-to-analog converter (DAC8568, TI, USA) on the mixed signal converter board is designed to configure the bias voltages of the CMOS chip. the mixed signal converter board is connected with the FPGA based control board through FPGA Mezzanine Card (FMC) connector. The Xilinx kintex-7 FPGA is the core logic control in a scalable and minimum hardware including an ethernet transceiver module, two Double-Data-Rate Three Synchronous Dynamic Random Access Memory (DDR3 SDRAM) module and a flash configuration module on the FPGA based control board. The FPGA based control board receives the serialized data output from the ADS5282 and transmits the data to PC through the ethernet.

The in-house built X-ray CT imaging benchtop has a medical-grade X-ray tube (G-242, Varex, UT, USA). The maximum tube voltage is 125 kVp, and the tube current is fixed at 12.5 mA during the continuous exposures. A beam collimator is installed to regulate the beam shape. The fixed beam filtration is 1.5 mm aluminum. The distance from the X-ray focal spot to the rotation center is 444.0~mm, and is 476.0~mm to the photosensitive layer of the detector. The horizontal linear stage (Model: LS12-X200, Hanjiang, China) has a travel distance of 200.0 mm, and the vertical linear stage (Model: STS06-X20, Hanjiang, China) has a travel distance of 40.0 mm. For CT imaging, the object is scanned on a rotation stage(Model: URS100BCC, Newport, USA) by 360 degree with 2.0 degree interval. The MTF and NPS experiments are measured with RQA3 beam, which is produced from a tube voltage of 50 kVp and filtered by 10.0 mm thick aluminum. Note that the CMOS detector is not placed 1500.0 mm away from the X-ray source as required by the IEC 62220-1 standard. The tungsten (W) plate has a thickness of 1.0 mm and is placed on the surface of the detector. The W plate blocks nearly half of the CMOS detector array. The CMOS detector array is positioned to align with the X-ray source focal spot when performing the NPS measurements. The acquired thousands of X-ray projection images from the CMOS were post-processed to stitch a full DR projection of the object. In particular, the signals correspond to the same spatial positions are combined together, and the signals belong to the different spatial positions are orderly rearranged.